\documentclass[final,3p,times]{elsarticle}

\usepackage{amssymb}

\usepackage{amssymb}
\usepackage[]{amsmath}
\usepackage{graphicx}
\usepackage{graphics}
\usepackage{subfigure}
\usepackage{latexsym}
\usepackage[]{amsmath}
\usepackage{amssymb}
\usepackage{indentfirst}
\usepackage{epsfig}
\usepackage{epstopdf}
\usepackage{float}
\usepackage{setspace}
\usepackage{booktabs}
\usepackage{times}
\usepackage{anysize}

\biboptions{numbers,sort&compress}
\usepackage[colorlinks, citecolor=red]{hyperref}
\usepackage{amsmath}
\usepackage{times}
\usepackage{anysize}
\marginsize{1.5cm}{1.5cm}{1cm}{1.5cm}
\selectfont

 \journal {}

\begin{document}

\begin{frontmatter}



\title{High order nonlocal symmetries and exact solutions of variable coefficient KdV equation}

\author{ Xiangpeng Xin \fnref{label2} $^*$  }

\cortext[cor1]{Corresponding author}
\ead{xinxiangpeng2012@gmail.com}

\author{Hanze Liu \fnref{label2} }
\author{Linlin Zhang \fnref{label2} }
\address[label2]{ School of Mathematical Sciences, Liaocheng University, Liaocheng 252059, PR China}

\begin{abstract}
In this paper, nonlocal symmetries and exact solutions of variable coefficient Korteweg-de Vries (KdV) equation are studied for the first time. Using pseudo-potential, high order nonlocal symmetries of time-dependent coefficient KdV equation are obtained. In order to construct new exact analytic solutions, new variables are introduced, which can transform nonlocal symmetries into Lie point symmetries. Furthermore, using the Lie point symmetries of closed system, we give symmetry reduction and some exact analytic solutions. For some interesting solutions, such as interaction solutions among solitons and other complicated waves are discussed in detail, and the corresponding images are given to illustrate their dynamic behavior.
\end{abstract}

\end{frontmatter}


\section{Introduction}

The theory of Lie group \cite{Lie1} was proposed by Norwegian mathematician Sophus Lie in the 19th century. In the 20th century, the Lie group theory developed rapidly\cite{Olver1,Ibragimov1,Bluman1}. Lie group theory was not only used to solve differential equations, but also established the relationships with many disciplines, such as nonlinear theory, integrable system, etc. Nowadays, Lie group theory has been widely applied to many fields, such as, mathematics, physics, numerical analysis, quantum mechanics, fluid dynamics system, etc.

As a generalization of the symmetry, a lot of studies have been devoted to seeking the generalized Lie point symmetry. P.J.Olver\cite{Olver1} construct a new type of symmetry by using recursion operator which was called nonlocal symmetry in the 80s of the last century. G.W. Bluman et al.\cite{Bluman2,Bluman4} presented many methods to find nonlocal symmetries of partial differential equations (PDEs)by using potential systems. F. Galas \cite{Galas1}obtained the nonlocal symmetries by using the pseudo-potentials of PDEs, and construct exact solutions by the obtained nonlocal symmetries firstly. Recently, Lou et al.\cite{Lou3,Hu2,Huang1,Xin1,Huang2,Xin2,Miao1} found that Painlev\'{e} analysis can also be used to construct nonlocal symmetries which was called residual symmetries.

Seeking nonlocal symmetry of nonlinear differential equations with variable coefficients is a meaningful and difficult task. Constructing high order nonlocal symmetry is also a very difficult work, because the higher order of symmetry can reflect the fundamental invariance of the equation. To construct high order nonlocal symmetry, the equation should have high order Lax pair, or high order pseudo-potential firstly. Many scholars have made researches on high order pseudo-potential. M.C. Nucci gave a way to construct the pseudo-potential of the nonlinear differential equation and some high order pseudo-potential of several nonlinear systems are given in \cite{Nucci1,Nucci2}. Here we consider the variable coefficient KdV equation, with the help of pseudo-potential, the high order nonlocal symmetries of this equation are obtained. Finally, by introducing new variables variables, the nonlocal symmetry is transformed into local symmetry and exact solutions are constructed by using the lie group theory.

This paper is arranged as follows: In Sec.2, the nonlocal symmetries were constructed by using the pseudopotential of variable coefficient KdV equation. In Sec.3, the process of transforming from nonlocal symmetries to local symmetries was introduced in detail. The finite symmetry transformation can be obtained by solving the initial value problem, and new exact solutions were constructed by using known solutions. In Sec.4 some symmetry reductions and exact solutions of the KdV equation were obtained by using the Lie point symmetry of closed system. Finally, some conclusions and discussions are given in Sec.5.

\section{Nonlocal symmetries of variable coefficient KdV equation}

 The time-dependent coefficient KdV equation\cite{Lou4} reads
\begin{equation}\label{ak-1}
u_t  = u_{xxx}  + 6uu_x  + Gu_x
\end{equation}
where $u = u(x,t)$  are the real functions, $G=G(t)$ is a real function of $t$. When $G=0$, Eq.(\ref{ak-1}) reduce to the well-known KdV equation, and this equation can be derived from the similarity reductions of the Kadomtsev-Petviashvili equation. Pseudo-potentials, Lax pairs and B\"{a}cklund transformations have been studied in \cite{Lou4}. The soliton solutions of (\ref{ak-1}) with time varying boundary conditions have been also studied by Chan and Li\cite{Chan1}using inverse scattering mathod. To our knowledge, nonlocal symmetries for Eq.(\ref{ak-1}) have not been obtained and discussed, which will be the goal of this paper.

The corresponding pseudo-potential has been obtained in\cite{Lou4},
\begin{equation}\label{ak-2}
\begin{array}{l}
 \phi _{xx}  =  - u\phi , \\
 \phi _t  = (G + 2u)\phi _x  - u_x \phi. \\
 \end{array}
\end{equation}

To seek the nonlocal symmetries of variable coefficient KdV equation(\ref{ak-1}), one must solve the following linearized equations,

\begin{equation}\label{ak-3}
\sigma _t^1  - \sigma _{xxx}^1  - 6\sigma ^1 u_x  - 6u\sigma _x^1  - \sigma ^2 u_x  - G\sigma _x^1  = 0,
\end{equation}
 $\sigma^1,\sigma^2$ are symmetries of $u$ and $G$, which means Eqs.(\ref{ak-1}) is form invariant under the transformations,
\begin{equation}\label{ak-3-1}
\begin{array}{l}
u \to u + \epsilon \sigma^1,\\
G \to G + \epsilon \sigma^2,\\
 \end{array}
\end{equation}
with the infinitesimal parameter $\epsilon $. Be different from Lie point symmetries, we assume nonlocal symmetries of the Eq.(\ref{ak-1})have the following form,
\begin{equation}\label{ak-4}
\begin{array}{l}
 \sigma ^1  = Xu_x  + Tu_t  - U, \\
 \sigma ^2  = TG_t  - \Delta , \\
 \end{array}
\end{equation}
where $X,T,U,\Delta$ are functions of $\{x,t,u,G,\phi ,\phi _x \}$. By substituting Eq.(\ref{ak-4}) into Eq.(\ref{ak-3}) and eliminating $u_t,\phi_{xx},\phi_{t}$ in terms of Eq.(\ref{ak-1}) and pseudo-potential Eqs.(\ref{ak-3}), it yields a system of determining equations for the functions $ X, T, U,\Delta$ , solving these determining equations can obtain,

\begin{equation}\label{ak-5}
\begin{array}{l}
 X = \frac{1}{3}x \tilde F_{1t}  +  F_2 ,~~~~T =\tilde F_1 ,~~~~U = \tilde c_1 u + \frac{{\tilde c_2 }}{2}\phi _x^2  + \tilde c_3 \phi \phi _x  + \tilde c_5 \phi ^2  + \tilde c_4 , \\
 \Delta  =  - \frac{1}{3}x\tilde F_{1tt}  - \frac{1}{3}(2G + 12u) \tilde F_{1t}  -  \tilde F_{2t}  - (3\tilde c_2 \phi ^2  + 6\tilde c_1 )u - 6\tilde c_5 \phi ^2  - 6\tilde c_4 , \\
 \end{array}
\end{equation}
where $\tilde c_i (i = 1,...,5)$ are five arbitrary constants and $ \tilde F_1,\tilde F_2$ are arbitrary functions of $t$.

\textbf{Remark 1:} It is show that the results(\ref{ak-5}) are high order nonlocal symmetries of variable coefficient KdV equation when $\tilde c_2$ or $\tilde c_3 \neq 0$, and they are local symmetries when $\tilde c_2=\tilde c_3=\tilde c_5=0$.

Nonlocal symmetries need to be transformed into local ones\cite{Lou3,Hu2} before construct exact solutions. Hence, We need to construct a new system called closed system, and Lie symmetries of the closed system contain the nonlocal symmetries obtained above.

\section{Localization of the nonlocal symmetry}

For simplicity, letting $\tilde c_1  = \tilde c_2  = \tilde c_4 =\tilde c_5   = 0,\tilde c_3  = 1, \tilde F_1(t) = \tilde F_2(t) = 0$ and introducing an auxiliary variable $\psi=\phi _x$ in formula (\ref{ak-5}) i.e.,
\begin{equation}\label{ak-6}
\begin{array}{l}
 \sigma ^1  =  - \phi \psi , \\
 \sigma ^2  = 0. \\
 \end{array}
\end{equation}

To localize the nonlocal symmetry (\ref{ak-6}), we have to solve the following linearized equations,
\begin{equation}\label{ak-8}
\begin{array}{l}
 \sigma _{x}^4  + \sigma ^1 \phi  + u\sigma ^3  = 0, \\
 \sigma _t^3  - \sigma ^2 \psi  - G\sigma _x^3  - 2\sigma ^1 \psi  - 2u\sigma _x^3  + \sigma _x^1 \phi  + u_x \sigma ^3  = 0, \\
  \sigma ^4=\sigma ^3_x\\
 \end{array}
\end{equation}
which is form invariant under the following transformation,
\begin{equation}\label{ak-3-1}
\begin{array}{l}
\phi \to \phi + \epsilon \sigma^3,\\
\psi \to \psi + \epsilon \sigma^4,\\
\end{array}
\end{equation}
with the infinitesimal parameter $\epsilon $, and $\sigma^1,\sigma^2$ given by (\ref{ak-6}). It is not difficult to verify that the solutions of (\ref{ak-8}) have the following forms,
\begin{equation}\label{ak-9}
\sigma ^3  = \phi f,~~~~\sigma ^4  = \phi_x f+ \phi f_x,
\end{equation}
where $f$ satisfies the following equations,
\begin{equation}\label{ak-9-1}
\begin{array}{l}
 f_x  = \frac{{\phi ^2 }}{4}, \\
 f_t  = \frac{{\phi ^2 G}}{4} - \frac{{\phi ^2 u}}{2} - \psi^2 , \\
 \end{array}
\end{equation}
it is easy to obtain the following result,
\begin{equation}\label{ak-9-2}
\sigma^5=f^2,
\end{equation}
which is form invariant under,
\begin{equation}\label{ak-3-1-1}
f \to f + \epsilon \sigma^5.\\
\end{equation}

One can see that the nonlocal symmetry (\ref{ak-6}) in the original space $\left\{ {x,t,u,G,\phi} \right\}$ has been successfully localized to a Lie point symmetry in the enlarged space $\left\{ {x,t,u,G,\phi,\psi,f} \right\}$. It is not difficult to verify that the auxiliary dependent variable $f$ just satisfies the Schwartzian form of the variable coefficient KdV equation,
\begin{equation}\label{ak-9-3}
\frac{{f_x }}{{f_t }} - \{ f;x\}  - G = 0
\end{equation}
where $\{ f;x\}  = \left( {\frac{{f_{xx} }}{{f_x }}} \right)_x  - \frac{1}{2}\left( {\frac{{f_{xx} }}{{f_x }}} \right)^2 $ is the Schwartzian derivative.

After we successfully transform the nonlocal symmetries(\ref{ak-6}) into local symmetries. New exact solutions can be constructed naturally by Lie group theory. With the Lie point symmetry(\ref{ak-6}),(\ref{ak-9}),(\ref{ak-9-2}), by solving the following initial value problem,
\begin{equation}\label{ak-9-4}
\begin{array}{l}
 \frac{{d\tilde u(\epsilon )}}{{d\epsilon }} = \tilde \phi (\epsilon )\tilde \psi (\epsilon ),~~~~~~~~~~~~~~~~~~~~~~\tilde u(0) = u, \\
 \frac{{d\tilde G(\epsilon )}}{{d\epsilon }} = 0,~~~~~~~~~~~~~~~~~~~~~~~~~~~~~~~~~\tilde G(0) = G, \\
 \frac{{d\tilde \phi (\epsilon )}}{{d\epsilon }} =  - \tilde \phi (\epsilon )\tilde f(\epsilon ),~~~~~~~~~~~~~~~~~~~\tilde \phi (0) = \phi , \\
 \frac{{d\tilde \psi (\epsilon )}}{{d\epsilon }} =  - \frac{1}{4}\tilde \phi ^3 (\epsilon ) - \tilde f(\epsilon )\tilde \psi (\epsilon ),~~~~\tilde \psi (0) = \psi , \\
 \frac{{d\tilde f(\epsilon )}}{{d\varepsilon }} =  - \tilde f^2(\epsilon )  ,~~~~~~~~~~~~~~~~~~~~~~~~~\tilde f(0) = f, \\
 \end{array}
\end{equation}
where $\epsilon$ is the group parameter, we arrive at the symmetry group theorem as follows:

\textbf{Theorem 1.} If $\left\{ {u,G,\phi ,\psi ,f} \right\}$ is the solution of the prolonged system (\ref{ak-1})(\ref{ak-2})and (\ref{ak-9-1}), so is $\left\{ {\tilde u,\tilde G, \tilde \phi, \tilde \psi,\tilde f}\right\}$

\begin{equation}\label{ak-9-5}
\begin{array}{l}
 \tilde u(\epsilon ) = \frac{{8\epsilon ^2 f^2 u - \epsilon ^2 \phi ^4  + 8\epsilon ^2 f\phi \psi  - 16\epsilon fu - 8\epsilon \phi \psi  + 8u}}{{8(\epsilon ^2 f^2  - 2\epsilon f + 1)}}, \tilde G(\epsilon ) = G,\\
\tilde \phi (\epsilon ) = \frac{\phi }{{1 - \epsilon f}},\tilde \psi (\epsilon ) = \frac{{4\epsilon f\psi  - \epsilon \phi ^3  - 4\psi }}{{4(\epsilon ^2 f^2  - 2\epsilon f + 1)}},\tilde f(\epsilon ) = \frac{f}{{1 - \epsilon f}}, \\
 \end{array}
\end{equation}
with $\epsilon$ is an arbitrary group parameter.

Here we give a simple example, starting from a trivial solution of (\ref{ak-1})
\begin{equation}\label{akk-1}
u=0,G=1,
\end{equation}
it's not difficult to derive the special solutions for the variables $\phi,\psi,f$ from(\ref{ak-2})and (\ref{ak-9-1}),
\begin{equation}\label{akk-2}
\phi  = \tilde c,\psi  = 0,f = \frac{{\tilde c^2 }}{4}(t + x).
\end{equation}

Using theorem 1, it's not hard to verify
\begin{center}
$u = \frac{{ - 2\tilde c^4 \epsilon ^2 }}{{\delta ^2 }},G = 1,\phi  = \frac{{4\tilde c}}{\delta },\psi  = \frac{{4\tilde c^3 \epsilon }}{{\delta ^2 }},f = \frac{{\tilde c^2 (t + x)}}{\delta },$
\end{center}
are still the solutions to the system(\ref{ak-1}),(\ref{ak-2})and (\ref{ak-9-1}),where $\delta  = \tilde c^2 \epsilon (t + x) - 4$, $\tilde c$ is an arbitrary constant. One can get more solutions by repeating the theorem 1. These solutions can not be obtained by traditional Lie group method, Darboux transformation mathod, etc. Therefore, a series of new exact solutions of (\ref{ak-1}) can be constructed.

To search for more similarity reductions and exact solutions of Eq.(\ref{ak-1}), we use classical Lie group method. Assume the symmetries of whole prolonged system have the vector form,
\begin{equation}\label{ak-9-6}
 V = \bar X\frac{\partial }{{\partial x}} + \bar T\frac{\partial }{{\partial t}} + \bar U\frac{\partial }{{\partial u}}+  \bar \Delta\frac{\partial }{{\partial G}}+\bar P\frac{\partial }{{\partial \phi}}+ \bar F \frac{\partial }{{\partial f }} ,
\end{equation}
where $\bar X,\bar T,\bar U,\bar \Delta,\bar P,\bar F $ are the functions with respect to ${ x,t,u,G,\phi,f } $, which means that the closed system is invariant under the transformations

\begin{equation}
 (x,t,u,v,G ,\phi,\psi,f )  \to (x + \epsilon \bar X,t + \epsilon \bar T,u + \epsilon \bar U,G+ \epsilon \bar \Delta, \phi + \epsilon \bar P,f + \epsilon \bar F),
\end{equation}
with a small parameter $\epsilon $. Symmetries in the vector form (\ref{ak-9-6}) can be assumed as
\begin{equation}\label{ak-10}
\begin{array}{l}
 \sigma ^1  = \bar Xu_x  + \bar Tu_t  - \bar U, \\
 \sigma ^2  = \bar TG_t  - \bar \Delta , \\
 \sigma ^3  = \bar X\phi _x  + \bar T\phi _t  - \bar P, \\
 \sigma ^4  = \bar Xf_x  + \bar Tf_t  - \bar F, \\
 \end{array}
\end{equation}
where $\bar X,\bar T,\bar U,\bar\Delta,\bar P,\bar F $ are the functions with respect to $\left\{ {x,t,u,G,\phi,\phi_x,f } \right\} $. And $\sigma ^i,(i=1,...,4)$ satisfy the linearized equations of the prolonged system, i.e.,
\begin{equation}\label{ak-10-1}
\begin{array}{l}
 \sigma _t^1  - \sigma _{xxx}^1  - 6\sigma ^1 u_x  - 6u\sigma _x^1  - \sigma ^2 u_x  - G\sigma _x^1  = 0, \\
 \sigma _{xx}^3  + \sigma ^1 \phi  + u\sigma ^3  = 0, \\
 \sigma _t^3  - \sigma ^2 \psi  - G\sigma _x^3  - 2\sigma ^1 \phi_x  - 2u\sigma _x^3  + \sigma _x^1 \phi  + u_x \sigma ^3  = 0, \\
 \sigma _x^4  - \frac{1}{2}\sigma ^3 \phi  = 0, \\
 \sigma _t^4  + \sigma ^3 \phi u + \frac{1}{2}\sigma ^1 \phi ^2  - \frac{1}{2}\sigma ^3 \phi G - \frac{1}{4}\sigma ^2 \phi ^2  + 2\sigma _x^3 \phi_x  = 0, \\
 \end{array}
\end{equation}

Substituting Eqs.(\ref{ak-10}) into Eqs.(\ref{ak-10-1}) and eliminating $u_t ,\phi_{xx},\phi_{t},f_x,f_t$ in terms of the closed system, determining equations for the functions $\bar X,\bar T,\bar U,\bar\Delta,\bar P,\bar Q,\bar F $ can be obtained, by solving these equations, one can get

\begin{equation}\label{ak-11}
\begin{array}{l}
 \bar X = \frac{{c_1 x}}{3} + \tilde F_3 ,\bar T = c_1 t + c_2 ,\bar U =  - \frac{{c_1 u}}{3} + c_3 \phi \phi _x , \bar \Delta  =  - \frac{{2c_1 G}}{3} - \tilde F_{3t} ,\\
\bar P =  - \phi (c_3 f - c_4 ),\bar F =  - c_3 f^2  + \frac{{(c_1  + 6c_4 )f}}{3} + c_5 , \\
 \end{array}
\end{equation}
where $c_i,(i=1,2,...,5)$ are arbitrary constants, $\tilde F_3=\tilde F_3(t)$ is arbitrary function of $t$.

\section{Symmetry reduction and exact solutions of variable coefficient KdV equation}

In this section, we will give two nontrivial similarity reductions and group invariant solutions of variable coefficient KdV equation(\ref{ak-1})under consideration $c_3 \ne 0$. Without loss of generality, we let $c_1=\tilde F_3  = 0$. By solving the following characteristic equation,
\begin{equation}\label{ak-13-1}
\frac{{dx}}{0} = \frac{{dt}}{{c_2 }} = \frac{{du}}{{c_3 \phi \phi _x }} = \frac{{dG}}{{c_2 }} = \frac{{d\phi }}{{c_4 \phi  - c_3 \phi f}} = \frac{{df}}{{c_5  + 2c_4 f - c_3 f^2 }},
\end{equation}
one can obtain
\begin{equation}\label{ak-14}
\begin{array}{l}
 G = C,f = \frac{{\alpha {\rm{tanh}}(\Theta ) + c_4 }}{{c_3 }},\phi  = F_2 (x)\sqrt {\Theta ^2  - 1} , \\
 u = \frac{1}{{2c_2 }}c_3 F_2^2 (x)F_{1x} (x)\tanh ^2 (\Theta ) + \frac{1}{{c_2 }}c_3 F_2 (x)F_{2x} (x)F_1 (x) -  \\
 ~~~~~~\frac{1}{\alpha }c_3 F_2 (x)F_{2x} (x){\rm{tanh}}(\Theta ) + F_3 (x), \\
 \end{array}
\end{equation}
where $\alpha  = \sqrt {c_3 c_5  + c_4^2 } ,\Theta  = \frac{{\alpha (F_1 (x) + t)}}{{c_2 }}$.

Substituting Eqs.(\ref{ak-14})into the prolonged system yields,
\begin{equation}\label{ak-19}
\begin{array}{l}
 F_2 (x) =  \pm \frac{{2\sqrt { - c_2 c_3 \alpha ^2 F_{1x} (x)} }}{{c_2 c_3 }}, \\
 F_3 (x) = \frac{1}{{4c_2^2 F_{1x}^2 }}\left( {8(c_3 c_5  + c_4^2 )F_1 F_{1x}^2 F_{1xx}  + 4(c_3 c_5  + c_4^2 )F_{1x}^4  + c_2^2 F_{1xx}^2  - 2c_2^2 F_{1x} F_{1xxx} } \right), \\
 \end{array}
\end{equation}
One can see that through the Eqs.(\ref{ak-14})and (\ref{ak-19}), if we know the form of $F_1(x)$, then $u$ can be obtained directly. We known that auxiliary dependent variable $f$ satisfies the Schwartzian form, by substituting $f = \frac{{\alpha {\rm{tanh}}(\Theta ) + c_4 }}{{c_3 }}$ into (\ref{ak-9-3}), one can get,
\begin{equation}
2c_2^2 FF_{xx}  - 3c_2^2 F_x^2  - 4(c_3 c_5  + c_4^2 )F^4  + 2c_2^2 c_6 F^2  - 2c_2^2 F = 0,
\end{equation}
where $F =F(x)=F_{1x }$.

It is not difficult to verify that the above equation is equivalent to the following elliptic equation,
\begin{equation}\label{ak-19-1}
F_x  = \frac{{\sqrt {4(c_3 c_5  + c_4^2 )F^4  + c_7 c_2^2 F^3  + 2c_6 c_2^2 F^2  - c_2^2 F} }}{{c_2 }}.
\end{equation}

It is know that the general solution of Eq.(\ref{ak-19-1}) can be written in terms of Jacobi elliptic functions. Hence, expression of solution (\ref{ak-14}) reflects the
wave interaction between the soliton and the Elliptic function periodic wave. A simple solution of Eq.(\ref{ak-19-1}) is given as,
\begin{equation}\label{ak-19-2}
F = a_0  + a_1 sn(x,n) + a_2 sn^2 (x,n),
\end{equation}
where $sn(x,n)$ is Jacobi elliptic function, substituting Eq.(\ref{ak-19-2}) into Eq.(\ref{ak-19-1}) yields following six solutions,
\begin{equation}\label{ak-19-3}
\begin{array}{l}
 \{ c_3  =  - \frac{{c_4^2 }}{{c_5 }},c_6  = 4 - 2n^2 ,a_0  = \frac{{ - 1}}{{4n^2  - 4}},a_1  = 0,a_2  = \frac{{n^2 }}{{4n^2  - 4}}\} ; \\
 \{ c_3  =  - \frac{{c_4^2 }}{{c_5 }},c_6  = 4n^2  - 2,a_0  = \frac{1}{{4n^2  - 4}},a_1  = 0,a_2  = \frac{{ - 1}}{{4n^2  - 4}}\} ; \\
 \{ c_3  = \frac{{c_2^2 (n^4  - 2n^2  + 1) - c_4^2 }}{{c_5 }},c_6  =  - \frac{{n^2 }}{2} + \frac{5}{2},a_0  =  - \frac{1}{{2n^2  - 2}},a_1  =  \pm \frac{n}{{2n^2  - 2}},a_2  = 0\} ; \\
 \{ c_3  = \frac{{c_2^2 (n^6  - 2n^4  + n^2 ) - c_4^2 }}{{c_5 }},c_6  = \frac{{5n^2 }}{2} - \frac{1}{2},a_0  =  - \frac{1}{{2n^2  - 2}},a_1  =  \pm \frac{1}{{2n^2  - 2}},a_2  = 0\} ; \\
 \end{array}
\end{equation}
with $c_2,c_4,c_5  \in R, 0 \le n \le 1$.

Substituting Eqs.(\ref{ak-19-3}),(\ref{ak-19-2}) and $F_{1x}=F$ into Eq.(\ref{ak-19}), one can obtain the solutions of $u$. Because the expression is too prolix, it is omitted here. In order to study the properties of these solutions of KdV equation, we give some pictures of $u$ as following,

\textbf{Case 1:} $a_2  = 0$


In Fig.1, we plot the interaction solutions between solitary waves and elliptic function waves expressed by $\{ c_3  = \frac{{c_2^2 (n^4  - 2n^2  + 1) - c_4^2 }}{{c_5 }},\\c_6  =  - \frac{{n^2 }}{2} + \frac{5}{2},a_0  =  - \frac{1}{{2n^2  - 2}},a_1  =  - \frac{n}{{2n^2  - 2}},a_2  = 0\}$  with parameters $c_2=c_3=c_5=1,c_4=0.1,n=0.8$. We can see that the component $u$ exhibits a soliton propagates on Jacobi elliptic sine function waves background for two cycles. In Fig.1, the first picture(a) shows that the height of the soliton is approximately 6 at $t=-10$. With the development of time, soliton produces elastic collisions with other waves, and the height of the soliton is changing continuously. Picture(e) shows that soliton reaches its highest height at $t=-2$. After the collision, the soliton reverts to the original height and continues to collide with the adjacent waves see the pictures($f \to i$). The corresponding 3d image is given which exhibits a soliton propagating on period waves background.

\textbf{Case 2:} $a_1 = 0$


In Figs.2, we plot another form of interaction solutions between solitary waves and elliptic function waves expressed by $\{ c_3  =  - \frac{{c_4^2 }}{{c_5 }},c_6  = 4n^2  - 2,a_0  = \frac{1}{{4n^2  - 4}},a_1  = 0,a_2  = \frac{{ - 1}}{{4n^2  - 4}}\}$  with parameters $c_2=c_3=c_4=c_5=1,n=0.8$. And the corresponding 3d image is given. For other types of solutions, we're not going to give their figures here.

\section{ Summary and Discussion}

In this paper, we have studied high order nonlocal symmetries and exact solutions of the variable coefficient KdV equation for the first time. First of all, starting from the known pseudo-potential of the variable coefficient equation, nonlocal symmetries are derived directly through a particular assumption. In order to take advantage of the nonlocal symmetries, auxiliary variables $\psi,f$ are introduced. Then, the primary nonlocal symmetry is equivalent to a Lie point symmetry of a prolonged system. Applying the Lie group theorem to these local symmetries, the corresponding group invariant solutions are derived. Secondly, several classes of exact solutions are provided in the paper, including some special forms of exact solutions. For example, exact interaction solutions among soliton and other complicated waves. In fact, it is of interest to study these types of solutions, for example, in describing localized states in optically refractive index gratings. In the ocean, there are some typical nonlinear waves such as the solitary waves and the cnoidal periodic waves.

It is very meaningful to study the nonlocal symmetries and exact solutions of variable coefficient integrable models. However, there is still a lot of work to be done. For example, in a large number of nonlocal symmetries of an integrable model which one can be localized. Is it possible to apply the nonlocal symmetry theory of constant coefficient differential equation to the variable coefficient differential equation? Above topics will be discussed in the future series research works.

\section{ Acknowledgments}

This work is supported by the National Natural Science Foundation of China (Nos. 11505090,1171041), Research Award Foundation for Outstanding Young Scientists of Shandong Province(No. BS2015SF009). The doctorial foundation of Liaocheng University under Grant No.318051413.

\small{

\end{document}
\begin{thebibliography}{99}
\bibitem{Lie1} S. Lie, \"{u}ber die Integration durch bestimmte Integrale von einer Classe linearer partieller Differential gleichungen, Arch Math, 6 (1881)328-368.
\bibitem{Olver1} P.J. Olver, Applications of Lie Groups to Differential Equations, Berlin: Springer, 1986.
\bibitem{Ibragimov1} N.H. Ibragimov, Transformation Groups Applied to Mathematical Physics, Boston, MA: Reidel, 1985.
\bibitem{Bluman1} G.W. Bluman, S.Kumei, Symmetries and Differential Equations, Springer-Verlag, New York 1989.



\bibitem{Bluman2} G.W. Bluman, A.F. Cheviakov and S.C. Anco, Applications of Symmetry Methods to Partial Differential Equations, Springer New York, 2010.
\bibitem{Bluman4} G.W. Bluman, A.F. Cheviakov, Framework for potential systems and nonlocal symmetries: Algorithmic approach, J Math. Phys., 46(2005) 123506.
\bibitem{Galas1} F. Galas, New nonlocal symmetries with pseudopotentials, J Phys A Math Gen, 25(1992) L981.
\bibitem{Lou3} S.Y. Lou, X.R. Hu, Y. Chen, Nonlocal symmetries related to B\"{a}cklund transformation and their applications. J Phys. A: Math. Theor. 45(2012) 155209.
\bibitem{Hu2} X.R. Hu, S.Y. Lou and Y. Chen, Explicit solutions from eigenfunction symmetry of the Korteweg-de Vries equation. Phys. Rev. E 85(2012) 85056607-1.
\bibitem{Huang1} L.L. Huang, Y. Chen, Nonlocal symmetry and similarity reductions for the Drinfeld-Sokolov-Satsuma-Hirota system,  Appl. Math. Lett., 64 (2017) 177-184.
\bibitem{Xin1} X.P. Xin, X.Q. Liu, Interaction Solutions for (1+1)-Dimensional Higher-Order Broer-Kaup System, Commun. Theor. Phys, 66(2016) 479-482.
\bibitem{Huang2} L.L. Huang, Y. Chen, Nonlocal symmetry and similarity reductions for a (2+1)-dimensional Korteweg-de Vries equation,  Nonlinear Dyn., 92 (2018) 221¨C234.
\bibitem{Xin2} X.P. Xin, Y.T. Liu, X.Q. Liu, Nonlocal symmetries, exact solutions and conservation laws of the coupled Hirota equations, Appl. Math. Lett., 55(2016) 63-71.
\bibitem{Miao1} Q. Miao, X.P. Xin, Y. Chen, Nonlocal symmetries and explicit solutions of the AKNS system. Appl. Math. Lett., 28(2014) 7-13.




\bibitem{Nucci1} M.C. Nucci, Pseudopotentials, Lax equations and B\"{a}cklund transformations for non-linear evolution equations, J. Phys. A: Math. Gen., 21(1988) 73-79.
\bibitem{Nucci2} M.C. Nucci, Riccati-type pseudopotentials and their applications,  Mathematics in science and engineering, 185(1992)399-436.


\bibitem{Lou4} S.Y. Lou, Pseudopotentials, Lax pairs and B\"{a}cklund transformations for some variable coefficient nonlinear equations, J. Phys. A: Math. Gen., 24 (1991) LS13-LS18.

\bibitem{Chan1} W. L. Chan, L.K. Shun, Nonpropagating solitons of the variable coefficient and nonisospectral Korteweg-de Vries equation,  J Math. Phys., 30(1989) 2521-2526.












\end{thebibliography}
